\documentclass[pra,aps,reprint,superscriptaddress,longbibliography,floatfix]{revtex4-2}

\usepackage{graphicx}
\usepackage[colorlinks=true,citecolor=blue,urlcolor=blue]{hyperref}
\usepackage[dvipsnames]{xcolor}

\usepackage{amsmath,amssymb,cases,bm}

\newcommand{\AB}[1]{\textcolor{gray}{#1}}

\begin{document}
\title{Dipole dynamics of an interacting bosonic mixture}

\author{Luca Cavicchioli}
\affiliation{Istituto Nazionale di Ottica, CNR-INO, 50019 Sesto Fiorentino, Italy}
\affiliation{\mbox{LENS and Dipartimento di Fisica e Astronomia, Universit\`{a} di Firenze, 50019 Sesto Fiorentino, Italy}}
\author{Chiara Fort}
\affiliation{Istituto Nazionale di Ottica, CNR-INO, 50019 Sesto Fiorentino, Italy}
\affiliation{\mbox{LENS and Dipartimento di Fisica e Astronomia, Universit\`{a} di Firenze, 50019 Sesto Fiorentino, Italy}}
\author{Michele Modugno}
\affiliation{Department of Physics, University of the Basque Country UPV/EHU, 48080 Bilbao, Spain}
\affiliation{IKERBASQUE, Basque Foundation for Science, 48013 Bilbao, Spain}
\affiliation{EHU Quantum Center, University of the Basque Country UPV/EHU, Leioa, Biscay, Spain}
\author{Francesco Minardi}
\affiliation{Istituto Nazionale di Ottica, CNR-INO, 50019 Sesto Fiorentino, Italy}
\affiliation{\mbox{LENS and Dipartimento di Fisica e Astronomia, Universit\`{a} di Firenze, 50019 Sesto Fiorentino, Italy}}
\affiliation{Dipartimento di Fisica e Astronomia, Universit\`{a} di Bologna, 40127 Bologna, Italy}
\author{Alessia Burchianti}\email{burchianti@lens.unifi.it}
\affiliation{Istituto Nazionale di Ottica, CNR-INO, 50019 Sesto Fiorentino, Italy}
\affiliation{\mbox{LENS and Dipartimento di Fisica e Astronomia, Universit\`{a} di Firenze, 50019 Sesto Fiorentino, Italy}}


\begin{abstract}
 We unravel the coupled dipole dynamics of a two-species Bose-Einstein condensate with tunable interspecies interaction. We produce a degenerate mixture of $^{41}$K-$^{87}$Rb in an optical trap and we study the dipole oscillations of both atomic species in the linear response regime. 
 Varying the interspecies interaction from the weakly to the strongly attractive side, we measure the frequencies and the composition of the two dipole eigenmodes. For enough strong interactions, even beyond the mean-field collapse, we find that the frequency of the low-energy eigenmode is determined only by the bare trap frequencies and the species population imbalance.
 The experimental results are well reproduced by numerical simulations based on two coupled Gross-Pitaevskii equations. Our findings provide a detailed picture of the dipole excitations in asymmetric bosonic mixtures. 
\end{abstract}

\maketitle

\section{Introduction}

Multicomponent quantum systems emerge in a variety of physical contexts ranging from condensed matter to cosmology. In multicomponent superfluids, the interactions among constituents give rise to a number of fascinating and still not fully understood phenomena, which may be relevant in superconductors \cite{Iskin206,Tajima_PRB_2019,Yerin2019,Salasnich_prb_2019}, liquid helium \cite{PhysRevLett.15.773, AB1975} and neutron stars \cite{1984ApJ,Andersson_2003, Kobyakov_2017}. Quantum degenerate atomic mixtures have proven to be ideal platforms to study multicomponent superfluidity \cite{Modugno_2002,Ferrier-Barbut_2014,Roy_2017}, thanks to the high control and tunability offered by ultracold gases experiments.

Superfluidity in mixtures drastically differs from the single-component case \cite{Mineev_1974,Law_2001} because of its critical dependence on the system's excitation spectrum. In the context of ultracold gases, the collective excitations of two-component superfluids confined in an external potential have been studied both theoretically \cite{Pu_1998, Ohberg_1999, Vidanovic_2013} and experimentally, with internal states \cite{Hall_1998, Maddaloni_2000, egorov_2013, Bienaime_2016} and with different atomic species \cite{Modugno_2002, Ferrier-Barbut_2014, Roy_2017, Wilson_2021}. 

The dipole mode is the most basic example of collective oscillations: in a single-component superfluid trapped in a harmonic potential, it corresponds to rigid oscillations of the density, whose frequency, equal the corresponding harmonic value, is independent of the interatomic interactions \cite{kohn1961}. For two interacting superfluids, this simple picture is no longer valid since the mutual interaction mixes the dipole modes of the two components and shifts their frequencies. Thus, in this case, the study of dipole dynamics allows to access the intercomponent interaction \cite{egorov_2013,Roy_2017} and may provide new insights into the nature of superfluidity \cite{Ferrier-Barbut_2014, Fava_2018}. The experimental realization of binary mixtures of Fermi-Bose superfluids \cite{Ferrier-Barbut_2014,Yao2016, Roy_2017} has renewed interest in the study of collective oscillations \cite{Miyakawa_2000, Arup_2007,maruyama2008, Bruun_PRA_2015, Wu_PRB_2018, Wen2019CollectiveOM}. Conversely, the excitation spectrum of Bose-Bose mixtures has remained relatively unexplored and the experiments have mainly focused on either symmetric spin mixtures \cite{Hall_1998, Maddaloni_2000, Hamner_2011, Zhang2012, Bienaime_2016}, strongly imbalanced \cite{Wilson_2021}, or immiscible two-species Bose-Einstein condensates (BECs) \cite{Modugno_2002}. Further, in the aforementioned works, no systematic analysis has been done varying the intercomponent interaction.
 
In this work, we fill the gap by studying the dipole modes of a tunable $^{41}$K-$^{87}$Rb bosonic mixture, confined in a harmonic trap. We induce the center-of-mass (CM) oscillations of both atomic species and, focusing on the linear response regime, we investigate their coupled dynamics. We measure the frequencies and the composition of the mixed dipole modes as a function of the interspecies interaction, from the weakly interacting to the strongly attractive regime, even beyond the mean-field collapse. For enough strong interactions, we observe that the two condensates almost move as one. In this regime, we directly explore the role of the population imbalance and find that the dipole dynamics is governed by the in-phase mode whose frequency depends on the bare trap frequencies and the population imbalance.
The experimental results are in good agreement with numerical simulations performed by solving two coupled Gross-Pitaevskii (GP) equations. We also discuss a simple model, based on the Ehrenfest theorem, which captures the main features of the dipole dynamics. 

\section{System preparation}

We prepare binary condensates of $^{41}$K and $^{87}$Rb, following the methods described in Refs.~\cite{DErrico2019, condmat5010021}. Briefly, we start by loading both atomic species in a hybrid potential, consisting of a magnetic quadrupole plus a dimple laser beam. Here, $^{87}$Rb is evaporatively cooled, while $^{41}$K is cooled by thermal contact with $^{87}$Rb. The cooling sequence ends with the production of a degenerate mixture, with both species in the $\left|F=1,m_{F}=1\right\rangle$ hyperfine ground-state, in a purely optical potential. The final trap is formed by the dimple beam and an auxiliary beam, crossing at 45$^\circ$ in the horizontal plane. The typical condensate atom number $N_{1}+N_{2}$ (hereafter we use the notation: $1 \rightarrow$ $^{41}$K and $2 \rightarrow$ $^{87}$Rb) is {$3 \times 10^5$} and the species population imbalance, $\alpha={N_1}/{N_2}$, is adjusted by controlling the number of $^{41}$K and $^{87}$Rb atoms that are initially loaded in the hybrid trap \footnote{The production of degenerate mixtures with $^{41}$K as the majority species is hindered by the ineffectiveness of sympathetic cooling observed once we decrease the starting number of $^{87}$Rb atoms.}.

The interspecies scattering length $a_{12}$ is magnetically tuned by the Feshbach field, an homogeneous magnetic field $B_{z}$, directed along the vertical $\hat{z}$ axis (aligned antiparallel to gravity). 
In the range where $B_{z}$ is varied, the intraspecies scattering lengths are both positive and almost constant: $a_{11} =62 a_0$ \footnote{A. Simoni, private communication.} and $a_{22} =100.4 a_0$ \cite{Marte2002}. We start the experiment with a non-interacting mixture by setting $B_z$ close to $72$~G, in correspondence of the zero-crossing point ($a_{12}=0$) between two interspecies Feshbach resonances \cite{Thalhammer2008}. 

The trapping potential minima of $^{41}$K, $z^0_{1}$, and $^{87}$Rb, $z^0_{2}$, are vertically shifted due to the different harmonic trap frequencies. 
In order to overlap the two minima, we compensate the differential sag with a vertical magnetic field gradient $b_z$, taking advantage of the different magnetic moments of $^{41}$K and $^{87}$Rb in the $B_z$ range \cite{condmat5010021}. The gradient is adiabatically ramped to $-16.6$~G/cm, ensuring $z^0_{1} \simeq z^0_{2}$. Then, $B_z$ is linearly varied in 20~ms from the zero-crossing to the target value of $a_{12}$ and, finally, the dipole dynamics is started by increasing $b_{z}$ up to $-14.6$~G/cm in about 1~ms. This changes the positions of the trapping minima along $\hat{z}$ by $\Delta_1 \simeq 2~\mu$m and $\Delta_2 \simeq 1~\mu$m, triggering the dipole dynamics. 

The trap shifts are chosen as a compromise between the contrasting requirements of small perturbations and an acceptable signal-to-noise ratio of the recorded oscillations. It is worth noticing that for the typical in-trap amplitude oscillations, used in this work, the maximum relative velocity between the two superfluids remains below the sum of their sound velocities. This represents a necessary condition for preventing the emergence of dissipation \cite{Law_2001,Castin_2015, Abad_2015}.

In the non-interacting case, the measured trapping frequencies are ($\omega_{x,1},\omega_{y,1}, \omega_{z,1}) \simeq 2\pi \times (65, 190, 185)$~Hz for $^{41}$K, and ($\omega_{x,2},\omega_{y,2},\omega_{z,2}) \simeq 2\pi \times (45, 140, 135)$~Hz for $^{87}$Rb. Since the trap frequencies are very sensitive to the alignment of the trap beams, they are recorded daily and used as a normalization factor for the dipole mode frequencies observed for $a_{12} \neq 0$. Once the dipole dynamics is excited, the two condensates are held for a variable time in the optical dipole potential, then they are released from the trap and let expand. 
After a time-of-flight (TOF) $t_{TOF}=$35 (38)~ms, we set $B_z=0$ and record the $^{41}$K ($^{87}$Rb) density distribution in the $yz$ plane by absorption imaging \footnote{The different TOF between $^{41}$K and $^{87}$Rb is due to the technical limitation of the CCD camera, once operated in the double-shot mode.}. The effect of the interspecies interactions during the expansion is minimized by extinguishing $b_{z}$ 1~ms later than the optical potential. The corresponding species-selective magnetic force contributes to separate the two falling condensates.
From the density distribution we extract the vertical displacements of the CM positions of $^{41}$K and $^{87}$Rb, $\Delta z_{CM,1}$ and $\Delta z_{CM,2}$.

From the size and the optical depth of the atomic clouds we determine the nominal atom numbers, that are subsequently multiplied by the calibration factors. The latter are obtained by matching the TOF expansion of single-species BECs with the expected theoretical behaviour \cite{Castin1996}. We estimate an uncertainty in the atomic number correction of 20\% and 10\% for $^{41}$K and for $^{87}$Rb, respectively.

\section{Coupled dipole oscillations}
\label{sec:Coupled}
As common to many physical situations, two coupled oscillators exhibit two oscillation modes whose frequencies differ from the natural frequencies of the uncoupled systems and whose amplitudes generally depend on the initial conditions. In our case, the two oscillating BECs are coupled by the coupling constant $g_{12}$ which is proportional to $a_{12}$: $g_{ij}={2\pi \hbar^{2} a_{ij}}/{m_{ij}}$, with 
$m_{ij}={m_{i}m_{j}}/{\left(m_{i}+m_{j}\right)}$ ($i,j=1,2$).

A simple theoretical description of the present system can be obtained by means of the Ehrenfest theorem, or, equivalently, of the sum-rules approach, as discussed in Appendix~\ref{sec:TM}. In this case the frequencies $\omega_{\pm}$ of the two modes have the analytical expression given by Eq.~(\ref{modefreq}), and their shifts from the bare trap frequencies ($\omega_{z,1}$ and $\omega_{z,2}$), for a given mixture, are fixed by $g_{12}$ and by the equilibrium density distributions.
In particular, in the attractive (repulsive) regime the eigenmode frequencies are raised (lowered) by the interspecies interactions. Analyzing the eigenvectors of the matrix (\ref{EhreMatrix}), one can gain information on the relative phase of the two modes \cite{fortissima}. More specifically, in the attractive regime, the low-energy mode corresponds to the in-phase oscillations of the two condensates, while the high-energy mode to the out-of-phase ones. The situation is reversed in the repulsive regime, where the in-phase dynamics occurs for the high-energy mode.

However, it is worth noticing that since this model relies on the knowledge of the equilibrium configuration only, its predictions may become inaccurate if during the dynamics the shape of density distributions deviates significantly from the initial one, (i.e., when the evolution of each condensate cannot be described by a rigid translation). Therefore, in order to have a more accurate description 
of the observed oscillations, we perform full dynamical GP simulations (see Appendix.~\ref{sec:TM}).
\begin{figure}[t!] 
\includegraphics[width=0.95\columnwidth]{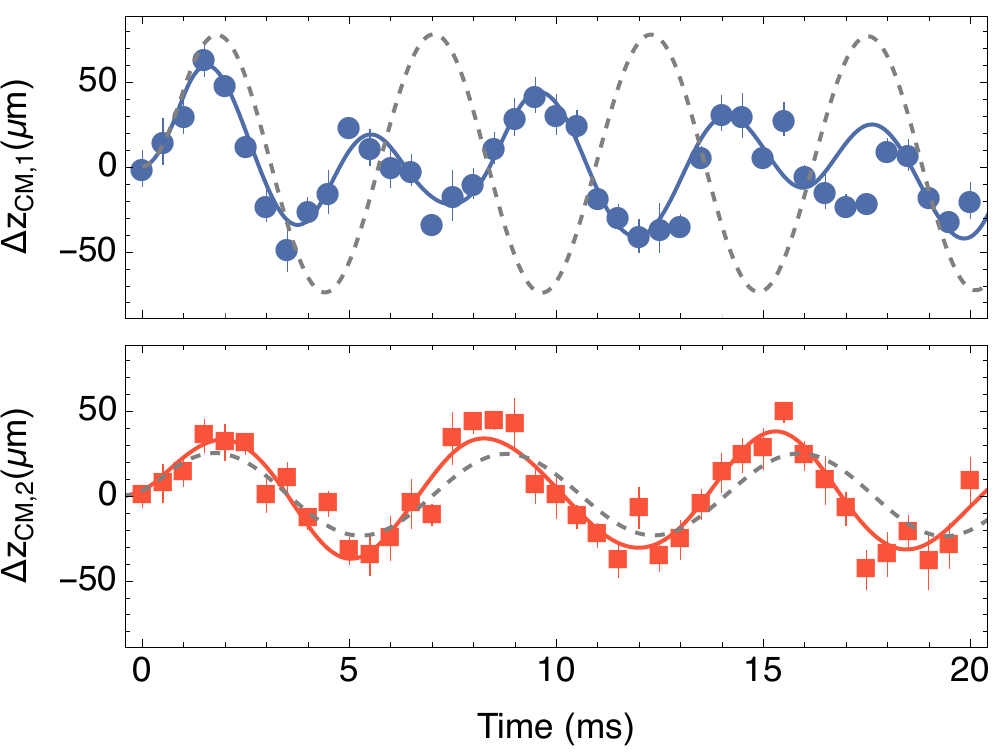}
\caption{\label{fig:fig1} Oscillations of $\Delta z_\mathrm{CM,1}$ (blue circle, upper panel) and $\Delta z_\mathrm{CM,2}$ (red squares, lower panel) in the weakly attractive regime, for $a_{12} =-(30.7 \pm 1.2) a_0$. The vertical CM positions, extracted from the density profiles, are measured after a TOF expansion of 35~ms (upper panel) and 38~ms (lower panel) for $^{41}$K and $^{87}$Rb, respectively. Error bars denote standard deviation of mean over five independent measurements. The continuous lines are the corresponding GP simulations where all parameters are fixed to the experimental ones ($\alpha=0.4$), with the exception of the initial phases, which are fitting parameters. As a reference, GP simulations in the non-interacting regime for the respective atomic species are also displayed as the grey dashed curves.}
\end{figure}

In Fig.~\ref{fig:fig1} we show the experimental CM shifts, $\Delta z_{CM,1}$ and $\Delta z_{CM,2}$, as a function of the in-trap time for a weakly attractive mixture with $a_{12} =-(30.7 \pm 1.2) a_0$, $N_1\simeq7\times 10^4$ and $\alpha \in [0.4,0.6]$. The variation range of $\alpha$ is due to fluctuations in the $^{41}$K and $^{87}$Rb atom numbers during the measurement. We observe that, in this regime, the $^{87}$Rb dipole dynamics is dominated by a single mode with frequency $\omega_-$, while the $^{41}$K one clearly shows two distinct modes with frequencies $\omega_-$ and $\omega_+$. The measurements are well reproduced by the corresponding GP simulations (continuous lines) where, by assuming that the interaction effects during the TOF are negligible, the expansion is obtained via the mapping $\Delta z_{CM}(t+t_{TOF})=\Delta z_{CM}(t) + t_{TOF}v_{CM}(t)$ ($v_{CM}$ being the vertical CM velocity at time $t$).
 In Fig.~\ref{fig:fig1} we also report, as a reference, the GP simulations in the non-interacting regime (dashed lines).
The comparison with the latter shows that the frequencies of both modes are up-shifted by the mean-field attraction, as expected (see Appendix \ref{sec:TM}).

\section{General Results and Discussion}
\label{sec:Results}
We have systematically investigated the dipole dynamics of the $^{41}$K-$^{87}$Rb mixture, varying the interaction coupling, from the weakly to the strongly attractive side. We restrict our analysis to this interaction range in order to remain in the linear response regime and to avoid the counter-flow instability \cite{Abad_2015}. Indeed, repulsive interactions increase the relative velocity between the two condensates by pushing them apart, and, at the same time, they are expected to lower the value of the critical velocity associated with the relative motion \cite{Law_2001}.

We measure both the frequencies ($\omega_-,\, \omega_+$) and the amplitudes, ($A_-,\,A_+$) \footnote{We remark that the frequencies $\omega_\pm$ are intrinsic properties of the system, whereas the amplitudes $A_\pm$ also depend on the protocol employed to excite the dynamics.}, of the dipole eigenmodes as a function of $a_{12}$. These quantities are extracted by fitting the CM positions of the two condensates with
\begin{equation}
\Delta z_ {CM}=\sum_{\kappa=\pm}A_{\kappa}e^{-t/\tau_{\kappa}}\sin(\omega_{\kappa} t+\phi_{\kappa}).
\end{equation}
The decay times ($\tau_{+},\, \tau_{-}$) are introduced to take into account damping effects which may emerge due to reasons both external and intrinsic to the dipole dynamics, as discussed in the following.

\begin{figure}[t!] 
\includegraphics[width=0.95\columnwidth]{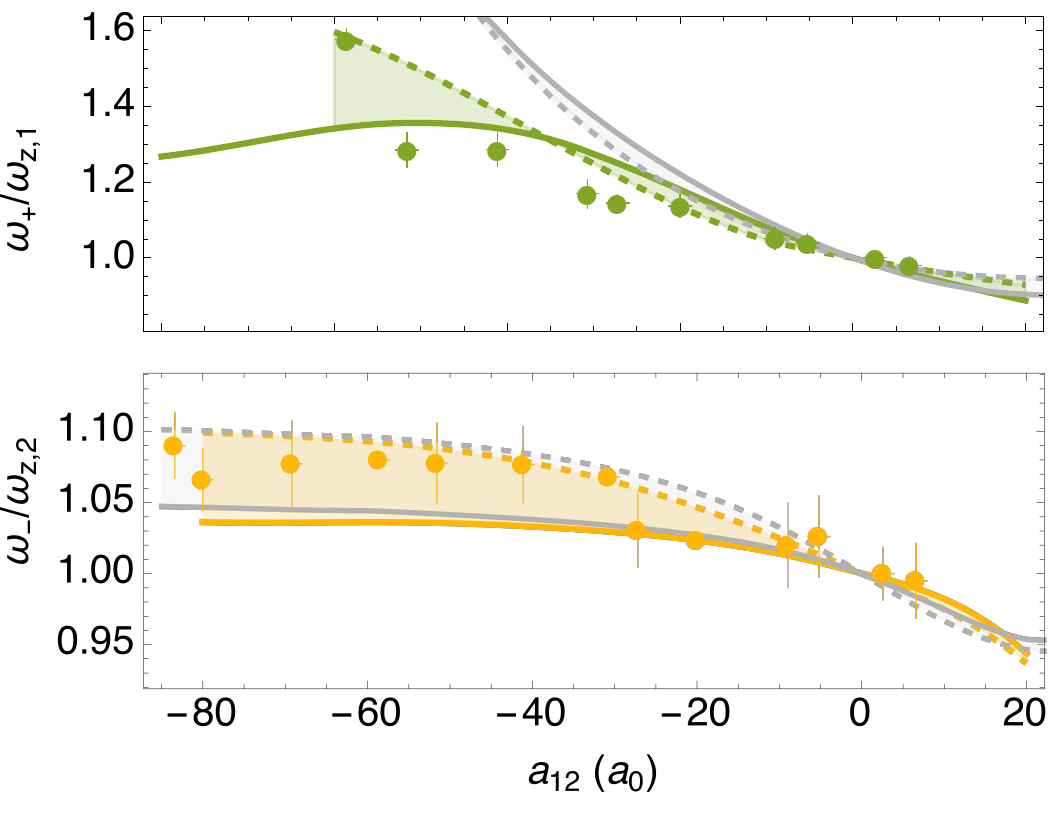}
\caption{\label{fig:fig2} Normalized eigenmode frequencies of a $^{41}$K-$^{87}$Rb mixture as a function of $a_{12}$: the frequency of the high-energy eigenmode is obtained by fitting the $^{41}$K oscillation (green circles, upper panel); the frequency of the low-energy eigenmode is obtained by fitting the $^{87}$Rb oscillation or, whenever possible (see text), it is given by the mean of the values obtained by fitting the motion of both atomic species (yellow circles, lower panel). The colored areas indicate the GP numerical predictions corresponding to the maximum experimental variation of $\alpha$: dashed colored lines  for $\alpha=0.84$ and continuous lines for $\alpha=0.32$. For comparison, the corresponding results from Eq.~(\ref{modefreq}) are also reported as gray lines: dashed for $\alpha=0.84$ and continuous for $\alpha=0.32$.}
\end{figure}

\begin{figure}[t!] 
\includegraphics[width=0.95\columnwidth]{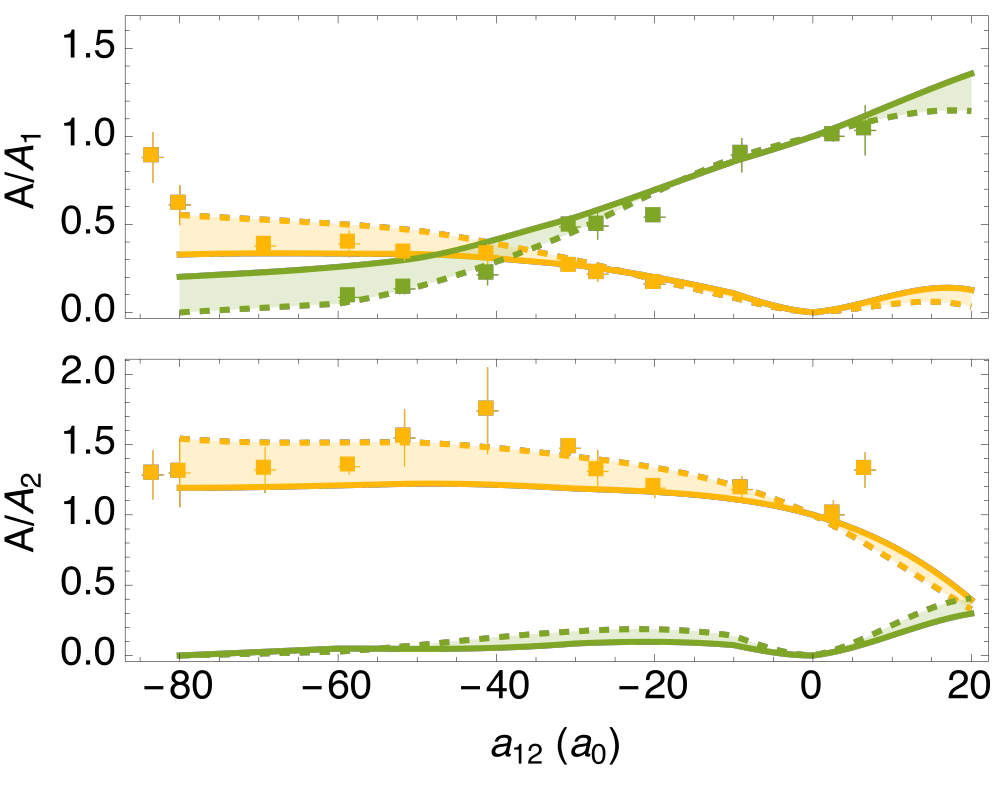}
\caption{\label{fig:fig3} Normalized eigenmode amplitudes of a $^{41}$K-$^{87}$Rb mixture as a function of $a_{12}$. Upper panel: experimental amplitudes of the low-energy (yellow squares) and high-energy (green squares) eigenmodes of $^{41}$K. 
Lower panel: experimental amplitude of the low-energy (yellow squares) eigenmode of $^{87}$Rb. 
The colored areas indicate the GP numerical predictions corresponding to the maximum variation of $\alpha$ during the measurements: the dashed line is for $\alpha=0.84$, while the continuous one is for $\alpha=0.32$.}
\end{figure}

We initially consider the case of slightly imbalanced mixtures, with $\alpha \in [0.32,0.84]$, $N_1\simeq7\times 10^4$ for $-80 a_0<a_{12}\leq 9 a_0$, and $N_1\simeq5\times 10^4$ for $a_{12}\leq -80 a_0$. 
In Fig.~\ref{fig:fig2}, we show the frequencies of the high and low-energy eigenmodes, respectively normalized to $\omega_{z,1}$ and $\omega_{z,2}$, versus $a_{12}$. On the attractive (repulsive) side, both $\omega_+$ and $\omega_-$ are up (down) shifted and this shift increases with the strength of the interspecies interaction.
The experimental data are compared to the results of the GP simulations (colored regions) and to the predictions of the Ehrenfest theorem in Eq.~(\ref{modefreq}) (gray regions).
In addition, in Fig.~\ref{fig:fig3} we show the amplitudes $A_+$ and $A_-$ extracted from the CM oscillations of both  $^{41}$K (upper panel) and $^{87}$Rb (lower panel). The amplitude of each condensate is normalized to the corresponding amplitude in the non-interacting case (that is, $A_1$ for $^{41}$K and $A_2$  for $^{87}$Rb).
In the weakly interacting regime, the $^{41}$K oscillations are characterized only by the high-energy eigenmode, with relative amplitude $A_+/A_1$ (data points). 
As the attractive interaction increases, also the low-energy eigenmode, with relative amplitude $A_-/A_1$ (data points), starts to appear and becomes dominant in the strongly attractive regime. On the contrary, the $^{87}$Rb oscillation is always dominated by the low-energy eigenmode and only the relative amplitude $A_-/A_2$ can be measured.

In order to reproduce the experimental results we perform corresponding GP simulations. The colored areas in Figs.~\ref{fig:fig2}-\ref{fig:fig3} show the numerical predictions obtained by considering the maximum variation of $\alpha$ during the measurements: the continuous line corresponds to $\alpha=0.32$ and the dashed one to $\alpha=0.84$.
The agreement between GP theory and experiment is good in the full range of explored interactions. 
For completeness, we also show in Fig.~\ref{fig:fig2} the eigenmode frequencies calculated with Eq.~(\ref{modefreq}). 
Approaching the strongly attractive side, we see that only the lower frequency mode, corresponding to the in-phase motion of the two condensates, is well reproduced by the Ehrenfest model.

Even for moderate couplings ($a_{12}<-10 a_0$), we measure a non-negligible damping of the high-energy dipole mode with a decay time $\tau_{+}$ of about 20~ms. This effect, observable also in the GP simulations, is the result of non-linear coupling to other collective modes for large enough excitations. We observe a decay of the dipole oscillations, depending on their amplitude, also for non-interacting mixtures due to the anharmonicity of the trap potential away from its minimum. However, in this case the decay time is of about 200~ms, for trap shifts of the order of 2~$\mu$m, substantially longer than $\tau_{+}$.

\begin{figure}[t!] 
\includegraphics[width=0.95\columnwidth]{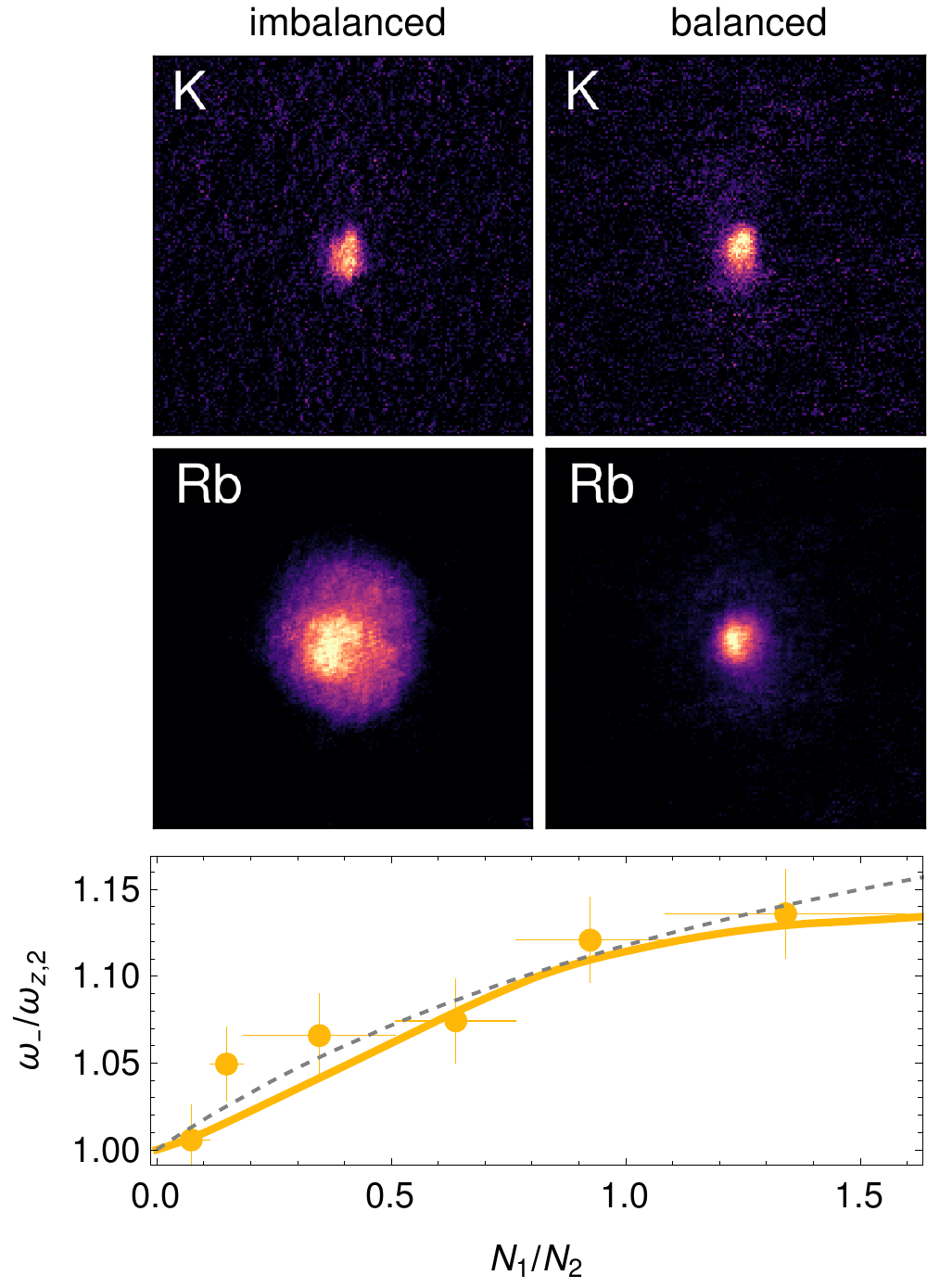}
\caption{ Upper panel: absorption images of a balanced and imbalanced dual-species BEC, in the strongly attractive regime, for $a_{12} =-(83.4 \pm 1.2) a_0$. Images are taken after 35~ms of TOF for $^{41}$K and 38~ms for $^{87}$Rb. 
$N_1= (4.8\pm1.0)\times 10^4$ and $N_2=(5.5\pm0.6)\times 10^4$, in the balanced case, and $N_1=(4.7\pm1.0)\times 10^4$ and $N_2=(3.4\pm0.3)\times 10^5$, in the imbalanced case. 
Lower panel: normalized frequency of the low-energy eigenmode as a function of the species population imbalance $\alpha$, for $a_{12}=-(83.4\pm 1.2) a_0$. The horizontal error bars correspond to the binning intervals of $\alpha$ values. The systematic uncertainty due to the atom number calibration mentioned in text corresponds to multiplying all $\alpha$ values for a scaling factor of $(1.0\pm 0.22)$. 
The continuous line is the result of GP simulations and the gray dashed line is the solution of Eq.~\ref{eq:omega}.}
\label{fig:fig4}
\end{figure}

For strong enough interactions, namely for $a_{12}<-70 a_0$, we observe that the two condensates oscillate in unison, with the same frequency and amplitude. In this limit, the eigenmode frequency $\omega_-$ tends to the value
\begin{equation}
\label{eq:omega}
\omega_-= \sqrt{\frac{  N_1 m_1 \omega_{z,1}^2+N_2 m_2 \omega_{z,2}^2}{N_1 m_1+ N_2 m_2}}.
\end{equation}
It is worth noticing that this expression depends only on the bare trap frequencies and on the total masses $N_i m_i$ of the two components.

We verify the accuracy of Eq.~(\ref{eq:omega}) in the strongly interacting regime for $a_{12}=-(83.4\pm 1.2) a_0$, where the mixture is stabilized against the mean-field collapse by quantum fluctuations \cite{Petrov}, i.e., for $g_{12} <- \sqrt{g_{11}g_{22}}$ (for our mixture the critical point corresponds to $a^{c}_{12}=-73.6 a_0$). To this end, we measure the low-energy eigenmode frequency once the species population imbalance spans a wider range, for $\alpha \in [0.07,1.34]$. 
In Fig.~\ref{fig:fig4}, we show the corresponding values of $\omega_-$ as a function of $\alpha$. The experimental data points are compared with the results of both the GP simulations including the Lee-Huang-Yang correction (continuous line, see Appendix~\ref{sec:TM}),
and Eq.~(\ref{eq:omega}) (dashed line). The frequency shift from the bare value $\omega_{z,2}$ decreases by decreasing $\alpha$ and eventually tends to $\omega_{z,2}$ when $\alpha \ll 1$.
 
Because the dipole dynamics is insensitive to the presence of liquid droplets, we probe their possible existence using TOF detection since, differently from the unbound gas component, droplets do not expand after the trap removal. Following the system evolution up to $t_{TOF}=25$ ms \footnote{For larger TOF the atoms exit, under the effect of gravity, the region where the Feshbach field is spatially homogeneous\cite{DErrico2019}}, we observe that, in the same conditions of Fig.~\ref{fig:fig4},  i.e., for $a_{12}=-83.4a_0$, both atomic species expand, even if the atom number exceeds the critical value for droplet formation  \cite{condmat5010021}. This effect can be ascribed to the fact that the system is prepared in a configuration far from the equilibrium one in free space. We verified that stable droplets form by further increasing the interspecies attraction. However, such increase implies a substantial reduction of the mixture lifetime,  due to the enhancement of three-body losses. This hinders the study of the in-trap dynamics in this range of parameters. We conclude that the production of long-lived droplet states requires a trap geometry matching the droplet size in free space, which is out the scope of the present work \cite{ferioli_dynamical,guo_lee-huang-yang_2021}.

\section{Conclusions}
We have studied the coupled dipole dynamics of a binary condensate by exploiting an asymmetric bosonic mixture, where the two components have different masses and experience different trapping potentials. In particular, we have investigated the dipole excitations of the two condensates as a function of their mutual coupling, exploring different regions of the mixture phase diagram. In the weakly interacting regime, where the mixture is miscible, we have excited both the in-phase and out-phase dipole modes, both observable in the CM oscillations of component $1$, and measured their dependence on the intercomponent interaction. We find that the frequencies of both eigenmodes are shifted by the coupling, differently from the case of symmetric spin mixtures, where the in-phase dipole frequency always matches the trap frequency. Further, we have extended our analysis to the strongly attractive regime, where the mixture is stabilized by quantum fluctuations. Approaching this region, the most significant features are the dominance of the in-phase dynamics, and the unison motion of the two condensates. In this limit, the dipole frequency is determined only by the bare trap frequencies and the total masses of the two components.
Our findings provide a solid ground for future experimental studies aimed at exploring the coupled dynamics of bosonic mixtures in different regimes, even beyond the linear limit, such as in the presence of dissipation or in proximity of a phase transition. Further extensions of our work could also include the exploration of higher-order excitations, which may be useful for probing quantum fluctuations \cite{LHY_fluid_Arlt}, exotic matter states \cite{PRA_Xia-Ji} and topological structures \cite{arxiv.shell} in multicomponent superfluids.

\begin{acknowledgments}

We thank Chiara d'Errico for early contributions to the experiment and the Quantum Gases group at LENS for support. We  acknowledge technical assistance of Antonio Orlando and the members of the LENS electronic and mechanical workshops. This work was supported by MUR Infrastructural funding through CNR-PASQUA initiative. MM acknowledges support by Grant PGC2018-101355-B-I00 funded by MCIN/AEI/10.13039/501100011033 and by “ERDF A way of making Europe”, and by the Basque Government through Grant No. IT1470-22.
\end{acknowledgments}
\appendix

\section{Theoretical Model}
\label{sec:TM}

The system under study can be described by two coupled GP equations of the form ($i,j=1,2$)
\begin{equation}
i\hbar\partial_{t}\psi_{i}(\bm{r},t)=\left[-\frac{\hbar^{2}}{2m_{i}}\nabla^{2}
+V^{tot}_{i}(\bm{r},t)\right]\psi_{i}(\bm{r},t),
\label{eq:GP}
\end{equation}
where $V^{tot}_i$ represents the total mean-field potential acting on the species $i$($\neq j$)
\begin{equation}
V^{tot}_{i}(\bm{r},t)= V^{ho}_{i}(\bm{r}) + g_{ii}n_{i}(\bm{r},t) + g_{ij}n_{j}(\bm{r},t),
\end{equation}
with $n_{i}(\bm{r},t)\equiv|\psi_i(\bm{r},t)|^2$ and $\int n_{i}=N_{i}$. 
The structure of the 
(collective) dipole modes along $z$
can be obtained by means of the Ehrenfest theorem \cite{ehrenfest}, namely 
\begin{equation}
\frac{d^{2}}{dt^2}\langle z\rangle_i=-\frac{1}{m_i}\langle \partial_{z}V^{tot}_i\rangle_i.
\end{equation}
 Then, by defining $\delta {z}_{i}(t)\equiv \langle {z}\rangle_{i}(t)$ it is straightforward to get \footnote{We have used the fact that $\int n_i\partial_{{z}}n_i=\int d(n_i^{2})/2\equiv0$ because of the vanishing boundary conditions.}
\begin{equation}
 \ddot {\delta {z}_{i}}(t) = - \omega^{2}_{i}\delta {z}_{i}(t) -\frac{g_{ij}}{m_{i}N_{i}}\int n_{i}\partial_{{z}}n_{j}.
 \label{eq:ehrenfest}
\end{equation}
In the \textit{limit of small oscillations}, the integral in the last term can be evaluated explicitly by assuming that the densities translate rigidly, $n_{i}({z},t)= n_{0i}[{z}-\delta {z}_{i}(t)]$ and use the following approximation: 
$n_{i}({z},t)\simeq n_{0i}({z}) -\partial_{{z}}n_{0i}({z}) \delta {z}_{i}(t)$ (for ease of notation, we have omitted the dependence on the transverse coordinates $x$ and $y$). 
Within these approximations, by defining $I\equiv\int\partial_zn_{01}\partial_zn_{02}$ and $\eta_{i}\equiv g_{12}I/(m_{i}N_{i})$, Eq. (\ref{eq:ehrenfest}) can be recast in matrix form as
\begin{equation}
\ddot{\delta {z}}= -M\delta {z},\quad\textrm{ with }
\delta {z}\equiv\begin{pmatrix}
 \delta {z}_{1} \\
 \delta {z}_{2}
 \end{pmatrix}
\end{equation}
and
\begin{equation}
M=\begin{pmatrix}
 \omega_{1}^{2} - \eta_{1} & \eta_{1}
 \\
 \eta_{2} & \omega_{2}^{2} - \eta_{2}
 \end{pmatrix}.
 \label{EhreMatrix}
\end{equation}

The frequencies $\omega_{\pm}$ of the two dipole modes are obtained from the eigenvalues of $M$, namely
\begin{align}
\label{modefreq}
 \omega^{2}_{\pm} &= \frac12\sum_{i=1}^2(\omega^{2}_{i} - \eta_{i}) 
 \\
 \nonumber&\pm \sqrt{\left[\sum_{i=1}^{2}(\omega^{2}_{i} - \eta_{i})/2\right]^{2}
 + \left(\omega_{1}^{2}\eta_{2} + \omega_{2}^{2}\eta_{1}\right) -\omega_{1}^{2}\omega_{2}^{2}}.
\end{align}
From the above equation one can easily
see that in the attractive regime ($g_{12}<0$) the two modes are up-shifted with respect to the bare trapping frequencies $\omega_1$ and $\omega_2$ of the two condensates, while in the repulsive ($g_{12}>0$) regime they are down-shifted.
It is also worth noticing that these frequencies are fixed by the equilibrium configuration of the system (i.e., the ground state in the same harmonic potential in which the evolution takes place) through the integral $I=\int\partial_zn_{01}\partial_zn_{02}$ \footnote{The equilibrium densities $n_{0i}(\bm{x})$ have to be computed numerically by minimizing the GP energy functional \cite{Pitaevski&Stringari} corresponding to Eq. (\ref{eq:GP}).}, as in the case of the sum-rules approach (see, e.g., Refs. \cite{Stringari_PRL_1996,Miyakawa_2000}). Indeed, it can be show that the two approaches are equivalent \cite{fortissima}.

For a general description, beyond the linear regime of small oscillations and of the rigid displacement ansatz [see below Eq. (\ref{eq:ehrenfest})], we perform dynamical simulations of the full GP equation (\ref{eq:GP}), to take into account the deformations of the density profiles during the evolution of the system. In addition, since at the mean-field level the mixture becomes unstable against collapse at $g_{12}=-\sqrt{g_{11} g_{22}}$, in the strongly attractive regime we also include the Lee-Huang-Yang (LHY) correction \cite{LeeHuangYang}, that accounts for the stabilizing effect of quantum fluctuations \cite{Petrov}. In particular, in Eq. (\ref{eq:GP}) we include a term $V_{\mathrm{LHY},i}\equiv \partial{\cal E}_{\mathrm{LHY}}/\partial n_{i}$, where the LHY energy density ${\cal E}_{\mathrm{LHY}}$ can be written in the form \AB{\cite{Ancilotto2018, Minardi2019}}
\begin{equation}
{\cal E} _{\rm LHY} = \frac{8}{15 \pi^2\hbar^3} 
 \left(g_{11}n_{1}m_{1}^{3/5} + g_{22}n_{2}m_{2}^{3/5}\right)^{5/2}.
\end{equation}

\bibliography{biblio}

\end{document}